\documentclass[11pt,a4paper,twoside,groupcitations]{article}
\usepackage[T1]{fontenc}
\usepackage[ansinew]{inputenc}
\usepackage[english]{babel}
\usepackage{amsfonts}
\usepackage{amsmath}
\usepackage{bm}
\usepackage{array}
\usepackage{amsthm}
\usepackage{wasysym}
\usepackage{amssymb}
\usepackage{graphicx}
\usepackage{subfigure}
\usepackage{bigints}
\usepackage{braket}
\usepackage{eucal}
\usepackage{verbatim}
\usepackage[table]{xcolor}
\usepackage{caption}
\usepackage{textcomp}
\usepackage{hyperref}
\usepackage{color}
\usepackage{mathtools}
\usepackage{commath}
\usepackage{bigints}
\usepackage[toc,page]{appendix}
\usepackage{csquotes}
\usepackage{multicol}
\usepackage{tikz}
\usetikzlibrary{positioning,arrows}
\usetikzlibrary{decorations.pathmorphing}
\usetikzlibrary{decorations.markings}
\usetikzlibrary{calc,decorations.markings}
\usetikzlibrary{arrows,shapes}
\usetikzlibrary{matrix,arrows}
\usepackage{pgfplots}
\usepackage{xparse}
\usepackage[export]{adjustbox}
\definecolor{jade}{HTML}{00A86B}
\newcommand{\be}{\begin{eqnarray}}
\newcommand{\ee}{\end{eqnarray}}
\newcommand{\pro}[2]{\mbox{$\langle\, #1 \mid #2\,\rangle$}}

\renewcommand{\d}{\mbox{${\rm d}$}} %d differenziale non corsivo in math mode
\newcommand{\lp}{\ell_{\rm p}}
\newcommand{\mpl}{m_{\rm p}}
\newcommand{\gn}{G_{\rm N}}

\newcommand{\Rh}{R_{\rm H}}

\graphicspath{{Figures/}}

\begin{document}
\title{\bf Mass (re)distribution for quantum dust cores\\ of black holes}
\author{L.~Gallerani$^{ad}$\thanks{E-mail: luca.gallerani6@unibo.it},
%$\ $
A.~Mentrelli$^{ade}$\thanks{E-mail: andrea.mentrelli@unibo.it},
%$\ $
A.~Giusti$^{bd}$\thanks{E-mail: A.Giusti@sussex.ac.uk},
and
R.~Casadio$^{cde}$\thanks{E-mail: casadio@bo.infn.it}
\\
\\
$^a${\em Department of Mathematics, University of Bologna}
\\
{\em Piazza di Porta San Donato 5, 40126 Bologna, Italy}
\\
\\
$^b$ {\em Department of Physics and Astronomy}
\\
{\em University of Sussex, Brighton, BN1~9QH, United Kingdom}
\\
\\
$^c${\em Department of Physics and Astronomy "A. Righi", University of Bologna}
\\
{\em via Irnerio~46, 40126 Bologna, Italy}
\\
\\
$^d${\em Alma Mater Research Center on Applied Mathematics (AM$^2$)}
\\
{\em Via Saragozza 8, 40123 Bologna, Italy}
\\
\\
$^e${\em I.N.F.N., Sezione di Bologna, I.S.~FLAG}
\\
{\em viale B.~Pichat~6/2, 40127 Bologna, Italy}
}
\maketitle
\begin{abstract}
The collective ground state for a spherical symmetric dust ball has been
investigated recently in R.~Casadio, \textit{Quantum dust cores of black holes},
Phys.~Lett.~B~\textbf{843} (2023) 138055.
In this study, we refine that model by obtaining a mass distribution
that accounts for the superposition of wavefunctions across different layers.
The refined mass distribution shows significant deviations from the approximation
without quantum superpositions.
Specifically, the new nearly parabolic distribution replaces the linear mass profile
of the original work, featuring an overall downward concavity, 
which leads to a non-vanishing tension.
%Finally, the validity of the findings is discussed in relation with the number of layers.
Notably, the regularity of the metric and causal structure are preserved in the refined
analysis.
\end{abstract}
\newpage
\section{Introduction}
\label{sec-1}
\setcounter{equation}{0}
General Relativity is the most successful (classical) theory of gravity to date.
Along its well known predictions, it also allows for the existence of black holes
with spacetime singularities hidden inside trapping surfaces~\cite{Penrose:1964wq}.
A way to mathematically resolve this issue could be to take the quantum nature of the
world into account.
The hope is that quantum physics may fix the singularity like it does for the hydrogen atom,
whose classical orbits would be unstable but simply do not exist in quantum mechanics
(see Refs.~\cite{Casadio:2019tfz, Haggard:2014rza} for a discussion of such approaches).
\par
In this article we propose to refine the quantum description of a dust ball recently introduced in
Ref.~\cite{Casadio:2023ymt}.
One of its main outcome is precisely the regularization of the Schwarzschild metric in
the interior $(0\leq r \leq\Rh)$, via an effective energy density $\rho$ that yields a
Misner-Sharp-Hernandez (MSH)~\cite{Misner:1964je,Hernandez:1966zia} mass 
\be
m(r)
\equiv
4\,\pi
\int_0^r
\rho(x)\,x^2\,\d x
\ ,
\label{m(r)}
\ee
which depends linearly on the areal radius $r$ inside the source.
To clarify this aspect let us briefly recall the main features of the model.
\par
The idea is to obtain a quantum state for the inner matter core of a spherically symmetric
black hole based on the Oppenheimer-Snyder model of dust collapse~\cite{Oppenheimer:1939ue}.
The construction starts by considering a perfectly isotropic ball of dust with total
Arnowitt-Deser-Misner (ADM)~\cite{Arnowitt:1959ah} mass $M$ and areal radius $r=R_{\rm s}(\tau)$,
where $\tau$ is the proper time measured by clocks comoving with the dust.
Dust particles are assumed to have the same proper mass $\mu\ll M$, and
will follow radial geodesics $r=r(\tau)$ in the Schwarzschild spacetime metric 
\be
\label{Schwarzschild}
\d s^{2}
=
-\left(1-\frac{2\,\gn\,m}{r}\right)
\d t^{2}
+\left(1-\frac{2\,\gn\,m}{r}\right)^{-1}
\d r^{2}
+r^{2}\,\d\Omega^2
\ ,
\label{schw}
\ee
where $m=m(r)$ is the (MSH) fraction of ADM mass inside the sphere of radius
$r=r(\tau)$~\footnote{We shall always use units with $c=1$ and often write the
Planck constant $\hbar=\lp\,\mpl$ and the Newton constant $\gn=\lp/\mpl$,
where $\lp$ and $\mpl$ are the Planck length and mass, respectively.}.
\par
We can discretise the ball by considering a spherical core of MSH mass $\mu_0=\epsilon_0\,M$
and radius $r=R_1(\tau)$, surrounded by $N$ comoving layers of inner radius $r=R_i(\tau)$,
thickness $\Delta R_i=R_{i+1}-R_i$, and mass $\mu_i=\epsilon_i\,M$, where $\epsilon_i$
is the fraction of ADM mass carried by the dust particles in the $i^{\rm th}$ layer.
The MSH mass in the ball $r<R_i$ will be denoted by 
\be
\label{Mcumul}
M_i
=
\sum_{j=0}^{i-1}\mu_j
=
M\,\sum_{j=0}^{i-1}\epsilon_j
\ ,
\ee
with $M_1=\mu_0$ and $M_{N+1}=M$. 
We also note that the radius $R_1$ and mass $M_1=\mu_0$ of the innermost core
as well as the thickness $\Delta R_i$ of each layer, can be made arbitrarily small by
increasing the number $N$ of layers in the classical picture.
The number $N$ should however remain such that the number of dust particles is
very large in each layer, a condition that will play an important role in the present
analysis.~\footnote{Ideally,
one would like to describe an astrophysical object of several solar masses which therefore
contains at least order of $10^{57}$ neutrons.}
\par
\begin{figure}[]
\centering
\begin{minipage}{.5\textwidth}
\centering
\includegraphics[width=\linewidth, height=0.25\textheight]{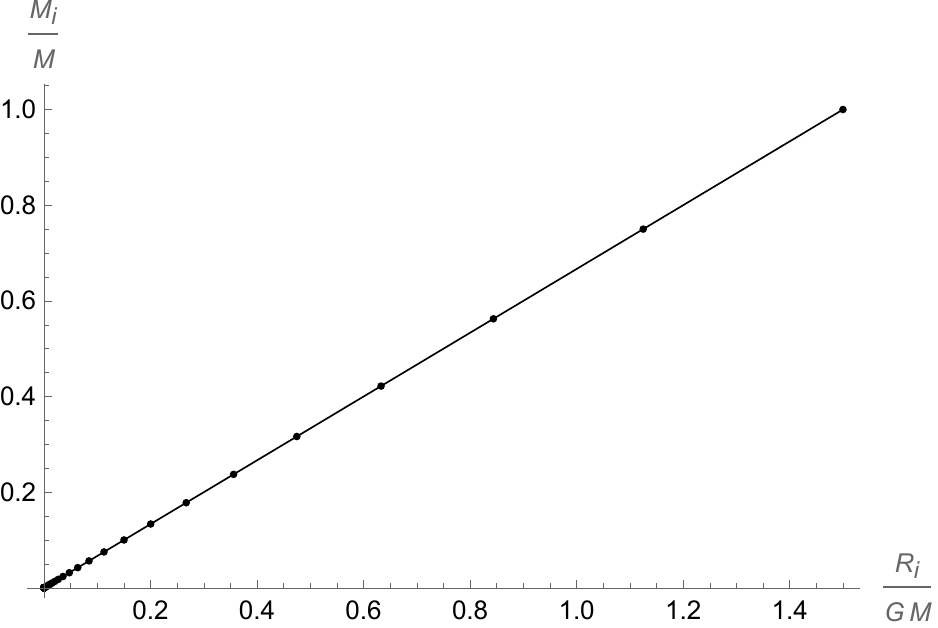}
\end{minipage}%
\begin{minipage}{.5\textwidth}
\centering
\includegraphics[width=\linewidth, height=0.25\textheight]{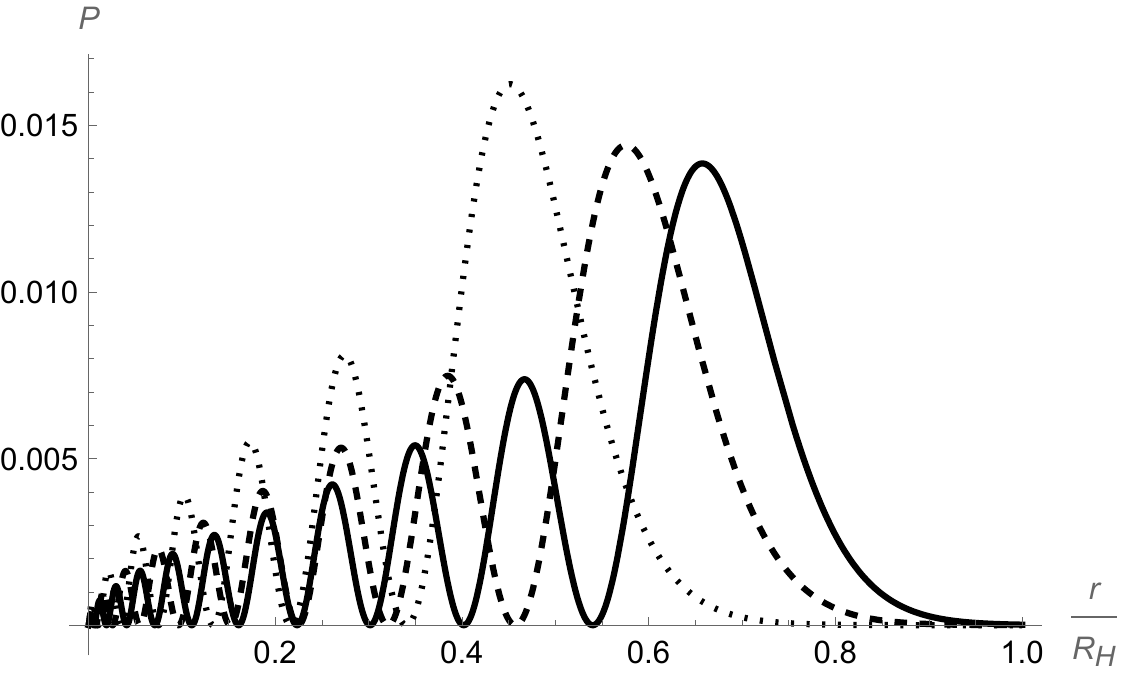}
\end{minipage}
\caption{Left panel: discrete mass function $M_i$ (dots) for $N=100$ layers
and its continuous approximation~\eqref{rho} (solid line).
Right panel: probability densities~\eqref{P} for $N=3$, $\mu = {\mpl}/{10}$, $M={(440/3)\,\mu}$.}
\label{N=3 & linear fit}
\end{figure}

The evolution of each layer can be derived by noting that dust particles located on the sphere
of radius $r=R_i(\tau)$ will follow the radial geodesic equation
\be
\label{geod-part}
H_i
\equiv
\frac{P_i^{2}}{2\,\mu}
-\frac{\gn\,\mu\,M_i}{R_i}
=
\frac{\mu}{2}\left(\frac{E_i^{2}}{\mu^2}-1\right)
%\equiv
%\mathcal E_i
\ ,
\ee
which defines the Hamiltonian $H_i$ for dust particles in the system, hence the layers they
are distributed on.
The canonical quantization prescription then leads to a time-independent Schr\"odinger equation,
whose solutions provide the Hamiltonian eigenstates $\ket{n_i}$.
The integral of motion $E_i$ for such states is well-defined only if the principal quantum number
is bounded below~\cite{Casadio:2021cbv} as
$n_i\geq N_i\equiv\mu\, M_i/m_\text{p}^2$~\cite{Casadio:2023ymt},
which therefore corresponds to the ground state wavefunction
\be
\label{psi}
\psi_{N_i}(R_i)
=
\sqrt{\frac{\mu\,\mpl}{\pi\,\lp^{3}\,M_i^{2}}}\,
\exp\!\left(-\frac{\mu\,R_i}{\mpl\,\lp}\right)
L_{\frac{\mu\,M_i}{\mpl^2}-1}^{1}\!\!
\left(\frac{2\,\mu\,R_i}{\mpl\,\lp}\right)
\ ,
\label{groundI}
\ee
where $L_{n-1}^1$ are generalized Laguerre polynomials and ${n}=1,2,\ldots$.
The states $\ket{n_i}$ are normalised in the scalar product which makes
$\hat H_i$ Hermitian, that is
\be
\pro{n_i}{n'_i}
=
4\,\pi\int_0^\infty R_i^2\,\psi_{n_i}^*(R_i)\,\psi_{n'_i}(R_i)\,\d R_i=
\delta_{n_i n'_i}
\ ,
\ee
so that the probability density to find a dust particle of the (inner surface of the) $i^{\rm th}$
layer at radial position $r$ is given by
\be
\label{P_i}
\mathcal{P}_i
=
4\,\pi\, r^2\, \abs{\psi_{n_i}(r)}^2
\ .
\label{P}
\ee
Examples for the ground states $n_i=N_i$ are plotted in the right panel of Fig.~\ref{N=3 & linear fit}.
\par
The expectation value of the areal radius on these ground states is given by~\footnote{To
simplify the notation, we replace the subscript $N_i$ with $i$ hereon.}
\be
\bar R_{i}\equiv\bra{N_i} \hat R_i \ket{N_i}
=
\frac{3\,\mpl^3\,\lp\,N_i^2}{2\,\mu^2\,M_i}
=
\frac{3}{2}\,\gn\,M_i
\ ,
\ee
with relative uncertainty
\be
\frac{\overline{\Delta R}_{i}}{\bar R_{i}}
\equiv
\frac{\sqrt{\bra{N_i}\hat R^2_i\ket{N_i}-\bar R_{i}^2}}{\bar R_{i}}
=
\frac{\sqrt{N_i^2+2}}{3\,N_i}
\simeq
\frac{1}{3}
\ ,
\label{DR/R}
\ee
which is used to determine the ground state thickness of the $i^{\rm th}$ layer.
From the construction of a global ground state under the assumptions that
$N_i$ is a large integer for all $i=1,\ldots,N$ and that the probability
density~\eqref{P} is negligible for $|r-\bar{R}_{i}|>\overline{\Delta R}_{i}$, 
one finds
\be
2\,\gn\,M_i
=
\frac{4}{3}\,\bar R_{i}
=
\bar R_{i}
+
\overline{\Delta R}_{i}
\simeq
\bar R_{i+1}
=
\frac{3}{2}\,\gn\,M_{i+1}
\ ,
\label{Miscaling}
\ee
or $M_{i+1}\simeq 4\,M_{i}/3$. 
The discrete mass function $M_i$ therefore grows linearly with the areal radius
$\bar R_{i}$ in the collective ground state, regardless of the number of layers $N$
we employ to describe it.
One can introduce a continuous effective energy density
\be
\rho
\simeq
\frac{M}{4\,\pi\,R_{\rm s}\,r^2}
\simeq
\frac{\mpl}{6\,\pi\,\lp\,r^2}
\ ,
\label{rho}
\ee
such that the effective MSH mass function~\eqref{m(r)} reads
\be
\label{mlinear}
m(r)
=
\frac{2\,\mpl\,r}{3\,\lp}
\label{MSH}
\ee
and equals the total ADM mass $M$ for $r=R_{\rm s}=\bar R_{N+1}$ (see left panel of
Fig.~\ref{N=3 & linear fit}).
Since dust particles in the ground state cannot collapse any further, the quantum core
is necessarily in equilibrium and the Schwarzschild geometry is replaced by an ``regular''
metric for $0\le r\le R_{\rm s}$.~\footnote{Technically, the core geometry corresponds to
an integrable singularity~\cite{Lukash:2013ts} in which no physical quantity diverges~\cite{Casadio:2024zek}.}
\par 
This summary illustrates how the model predicts a linear relation between mass and radius,
as a first approximation.
However, the profile of $\mathcal{P}_i$ shown in the right panel of Fig.~\ref{N=3 & linear fit},
suggests that particles classically belonging to the $i^{\text{th}}$ layer have a non-vanishing
probability to be localised elsewhere.
This mechanism should affect the mass of every layer, leading to a new MSH mass distribution
$\mathcal{M}_i$, that will likely not preserve linearity.
This effect can be seen as a first order correction on top of the linear behaviour
described above, and its detailed derivation is given in Section~\ref{sec2}
(an alternative but equivalent formulation is also provided in Appendix~\ref{A.1}).
Section~\ref{sec3} then compares the new results with those previously reported
in Ref.~\cite{Casadio:2023ymt}.
Conclusions from the present analysis are drawn in Section~\ref{sec-4}, along with an
outlook on future developments.
\section{Refined mass distribution with quantum superpositions}
\label{sec2}
\setcounter{equation}{0}
We want to estimate the correction to the linear relation~\eqref{MSH} following
from the radial profile of the ground state wavefunctions~\eqref{groundI}.
From Eq.~\eqref{Mcumul}, the contribution to the MSH mass from the
$i^{\rm th}$ layer is defined by
\be
\mu_{i}
=
M_{i+1}-M_i
\ ,
\qquad 
\forall\, 
i=1,\ldots,N
\ .
\ee
Using the approximate result in Eq.~\eqref{Miscaling}, we find
\be
\mu_i
=
\frac{M_i}{3}
=
\frac{1}{4}\left(\frac{3}{4}\right)^{N-i}
M
\ ,
\label{mu_i}
\ee
being $M \equiv M_{N+1}$ the total ADM mass.
%This approximation holds provided $\mu_0=M_1\gg \mu$.
Since the probability density $\mathcal{P}_i$ in Eq.~\eqref{P} does not vanish for
$|r-{R}_{i}|>{\Delta R}_{i}$,~\footnote{We omit the bar over expectation values for simplicity,
so that $\bar R_i=R_i$ from now on.}
the probability that particles belonging to the $i^{\rm th}$ layer are actually found inside
a different layer is not zero, which affects the mass of both layers.
The argument extends to all particles in the ball, so that one expects
that the actual contribution to the MSH mass will differ from the expression~\eqref{mu_i}
and the distribution $M_i$ will differ from the linear behaviour~\eqref{Miscaling}.
\par
An efficient way to keep track of the mass contributions that each $\psi_i$ brings
in all the other layers is to construct an $(N+1)\times (N+1)$ matrix $\Delta \mu$,
whose entries $\Delta\mu_{i,j}$ are the contributions to the fraction of MSH mass
$\mu_{j-1}$ inside the $(j-1)^{\rm th}$ layer coming from dust particles in the ground
state of the $(i-1)^{\rm th}$ layer, that is
\be
\label{Delta mu}
\Delta{\mu}_{i,j}
=
\mu_{i}\int_{R_{j}}^{R_{j+1}}\mathcal{P}_{i}(r)\, \d r
\ ,
\quad \quad
i,j=0,\ldots,N
\ ,
\ee
where $R_0=0$~\footnote{We identify with $R_{(j=0)}=0$ the origin of the ball.}
and $\mathcal{P}_0$ is the probability density for dust particles
in the innermost core.
\par
It is easy now to construct the matrix $\Delta \mu$ as follows:
\be
\label{deltamu}
\begin{split}
\Delta \mu
=
&
\begin{pmatrix}
\mu_0\bigintss_{0}^{R_1}\mathcal{P}_0 \, \d r
&
\mu_0\bigintss_{R_1}^{R_2}\mathcal{P}_0 \, \d r
& 
\ldots
&
\mu_0\bigintss_{R_N}^{R_{\rm s}}\mathcal{P}_0 \, \d r
\\[6pt]
\mu_1\bigintss_{0}^{R_1}\mathcal{P}_1 \, \d r
&
\mu_1\bigintss_{R_1}^{R_2}\mathcal{P}_1 \, \d r
&
\ldots
&  \mu_1\bigintss_{R_N}^{R_{\rm s}}\mathcal{P}_1 \, \d r
\\[6pt]
\ldots & \ldots & \ldots & \ldots \\[6pt]
\mu_N\bigintss_{0}^{R_1}\mathcal{P}_{N} \, \d r
&
\mu_N\bigintss_{R_1}^{R_2}\mathcal{P}_{N} \, \d r
&
\ldots
&
\mu_N\bigintss_{R_N}^{R_{\rm s}}\mathcal{P}_{N} \, \d r
\end{pmatrix}
\\
=
& 
\begin{pmatrix}
\mu_0 & 0 & \ldots & 0 \\[6pt]
\Delta{\mu}_{1,0} & \Delta{\mu}_{1,1} & \ldots & \Delta{\mu}_{1,N} \\[6pt]
\ldots &\ldots &\ldots&\ldots \\[6pt]
\Delta{\mu}_{N,0} & \Delta{\mu}_{N,1} & \ldots & \Delta{\mu}_{N,N} \\[6pt]
\end{pmatrix}
\ ,
\end{split}  
\ee
where we assumed that dust particles in the innermost core have negligible probability
to leak into larger layers and all $\mu_i$ are given in Eq.~\eqref{mu_i}.
The reason for this assumption is that the innermost core can be made arbitrarily
small, and can thus contain a number of dust particles which is negligible compared to the rest of 
the layers.
This small number of particles would then not produce a significant effect at larger 
radii, also given that their probability density is exponentially suppressed for increasing $r$,
as discussed above.
\par
We note that the probabilities $\mathcal{P}_i$ vanish for $r\to\infty$, and the probability
of finding dust particles in the range $[R_{\rm s}=R_{N+1},\infty)$ is not zero.
However, this tail of probability is essentially negligible for very large $M\gg \mu$, and
we can formally take care of it in the calculation by setting $R_{N+1}\to \infty$.
\par
The redistributed mass in the $j^{\rm th}$ layer is obtained by summing the terms
in the $\left(j+1\right)^{\rm th}$ column of the matrix $\Delta\mu$, i.e. 
\be
\Delta{\mu}_j
=
\Delta{\mu}_{0,j}+\Delta{\mu}_{1,j}+\ldots +\Delta{\mu}_{N, j}
=
\sum_{i=0}^{N} \Delta \mu_{i,j}\ , \qquad j=0,\ldots, N.
\ee
Similarly to Eq.~\eqref{Mcumul}, we finally define the MSH mass function so obtained as
\be
\label{finalM}
\mathcal{M}_i
=
\sum_{j=0}^{i}
\Delta{\mu}_j
\ ,
\qquad
i=0,\ldots,N
\ .
\ee
The above expression can be computed numerically and then compared to the initial linear 
distribution $M_i$ in Eq.~\eqref{Miscaling}.
This comparison will be carried out in the next Section.
Before that, we conclude with a final consideration regarding the validity
of our construction in relation to the number of layers $N$.
Notice that the value of $\mathcal{M}_i$ depends on the initial choice of $M$,
which fixes $M_i$, $\mu_i$ and $\mathcal{P}_i$.
In particular, $\mathcal{P}_i$ is determined also by the value of
\be
\label{Ni}
N_i
\simeq
\left(
\frac{3}{4}
\right)^{N-i+1}
\frac{\mu \, M}{m^2_p}
\ee
that represents both the quantum number for the $i^{\text{th}}$ wavefunction $\psi_i$,
and the order of its generalised Laguerre polynomial.
Relation~\eqref{Ni} reveals a degeneracy $N_i=1$ for a large number of layers
as a consequence of the integer nature of $N_i$, i.e. for a given $M$,~\footnote{The
analysis does not depends on the proper mass $\mu$ which, in practice, is defined
by the particle type.}
the relation~\eqref{Ni} may fail to capture the differences
among the $N_i$ when the integer $N-i$ is large.
For instance, when  $M=3000\,\mpl$ with $\mu=\mpl/10$ and $N=80$, one obtains
 \be
 \label{N80}
 N_i=
 (
 \underbrace
 {1,\ldots \, ,1}
 _
 {N=60},
 \underbrace
 {2,\, \ldots \,,\, 300}
 _
 {N=20}
 ) \, ,
 \ee
 while for the same values of masses, but $N=100$,
 \be
 \label{N100}
 N_i=
 (
 \underbrace
 {1,\ldots\, ,1}
 _
 {N=80},
 \underbrace
 {2,\, \ldots\,,\, 300}
 _
 {N=20}
 )
 \ .
 \ee
The conclusion is that only specific relations between $N$ and $M$ yield strictly
monotonic sequences of $N_i$, which remain unaffected when $N$ increases
and do not worsen the degeneracy.
In the approximation of Ref.~\cite{Casadio:2023ymt}, adding more layers to a certain
value of $N$ simply adds points near the core, that still satisfy the linear
relation~\eqref{Miscaling}, as shown in the left panel of Fig.~\ref{plot dots}.
However, all the red dots in that graph correspond to $N_i=1$ and their addition
to the system does not affect the outer layers, as the above example from~\eqref{N80}
to~\eqref{N100} suggests.
In fact, when $N=80$, the model predicts $N+1=81$ pairs of masses and radii,
one for each layer, which can be ordered from the smallest to the largest in the
sequence 
\be
\label{80layers}
\left\{(R_1,\mu_0)_{\vert_{N=80}},\
(R_2,\mu_1)_{\vert_{N=80}},
\ldots,
(R_{81},\mu_{80})_{\vert_{N=80}}
\right\}
\ .
\ee
Increasing $N$ to $100$ for the same $M$, is equivalent to adding $20$ pairs
in front of those in~\eqref{80layers}, while leaving the outer $81$ unchanged,
\be
\label{100layers}
\begin{split}
\left\{
(R_1,\mu_0)_{\vert_{N=100}},
\ldots,
(R_{101},\mu_{100})_{\vert_{N=100}}
\right\}
=
&\
\left\{
(R_1,\mu_0)_{\vert_{N=100}},
\ldots,
(R_{20},\mu_{19})_{\vert_{N=100}},
\right.
\\
&\
\left.
\quad
(R_1,\mu_0)_{\vert_{N=80}},
\ldots,
(R_{81},\mu_{80})_{\vert_{N=80}}
\right\}
\ .
\end{split}
\ee
This pattern suggests that the physics of the system is reasonably
captured by the outer layers carrying the higher factions of MSH mass.
\par
Similarly, when quantum superpositions are accounted for, increasing $N$ for
fixed $M$, corresponds to adding layers near the core.
However, in the original approximation, introducing layers characterised by $N_i=1$,
does not seem to spoil the linear relation~\eqref{Miscaling} between masses and radii.
In the presently refined version, such degeneracies should instead be avoided
as they are not accurately accounted for in Eq.~\eqref{Ni}, where for small $i$,
slightly different $N_i$ are treated as equal integers, leading to unphysical values
of $\mathcal{P}_i$.
To fully appreciate the contribution of the mass correction, it is essential to work
with pairs of values for $N$ and $M$ that ensure an increasing monotonic
sequence of the $N_i$.
This can be achieved either by reducing the number $N$ of layers or by increasing
the value of the total mass for $N$ fixed.
While both approaches are theoretically valid, in practice, large values of $M$
become computationally very demanding.
For instance, for $M \sim 10^5\,\mpl$ we could only find $N\sim 30$ layers
that ensure $N_i\gg 1$.
For this reason, in the next Section, we will opt for a lower value of $N$
to ease numerical evaluations.
\section{Effective metric and energy-momentum tensor}
\label{sec3}
\begin{figure}[]
\centering
\begin{minipage}{.5\textwidth}
\centering
\includegraphics[width=\linewidth, height=0.25\textheight]
{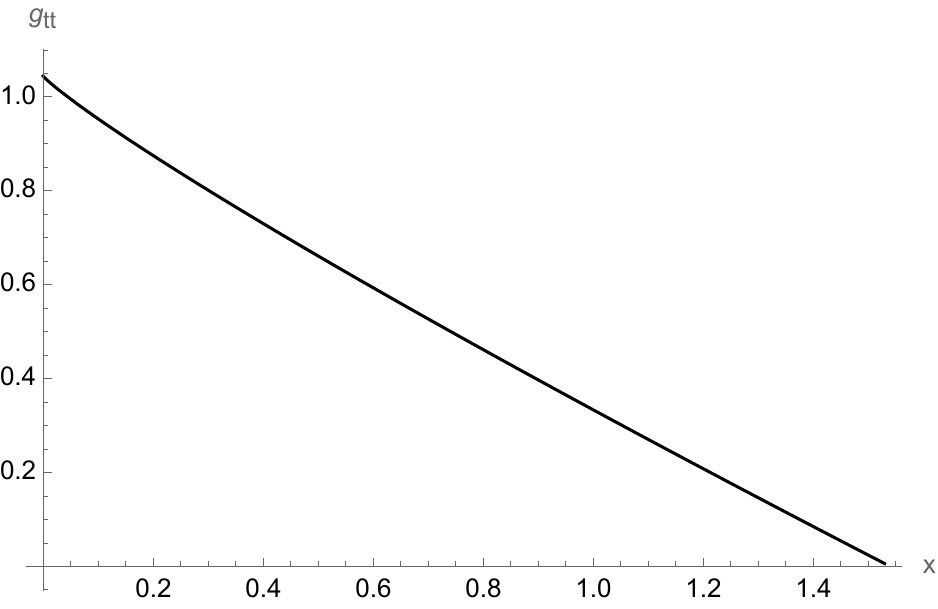}
\end{minipage}%
\begin{minipage}{.5\textwidth}
\centering
\includegraphics[width=\linewidth, height=0.25\textheight]
{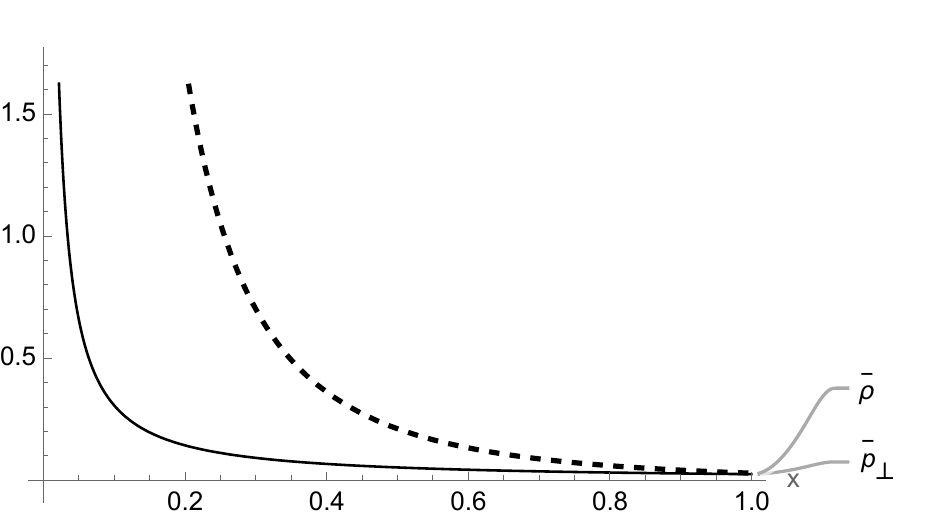}
\end{minipage}
\caption{Left panel: $g_{tt}$ for the metric~\eqref{gregular} (its zero lies outside $R_{\rm s}$,
at $x>1$).
Right panel: normalized $\bar{p}_{\perp}=R^2_{\rm s} \, \gn\,p_{\perp}$ (continuous line)
and $\bar{\rho}=R^2_{\rm s} \, \gn\,\rho$ (dashed line).
Both panels are obtained for $a=1.53$, $b=-0.533$, and $c=1.90$.}
\label{gtt}
\end{figure}

\setcounter{equation}{0}
\begin{figure}[]
\centering
\begin{minipage}{.5\textwidth}
\centering
\includegraphics[width=\linewidth, height=0.25\textheight]{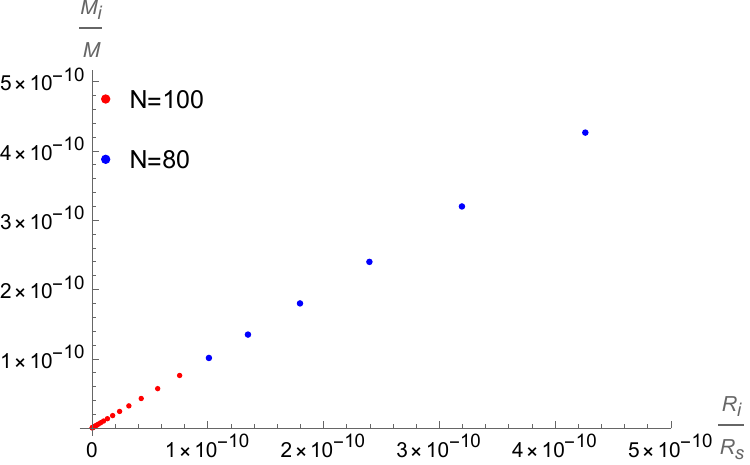}
\end{minipage}%
\begin{minipage}{.5\textwidth}
\centering
\includegraphics[width=\linewidth, height=0.25\textheight]{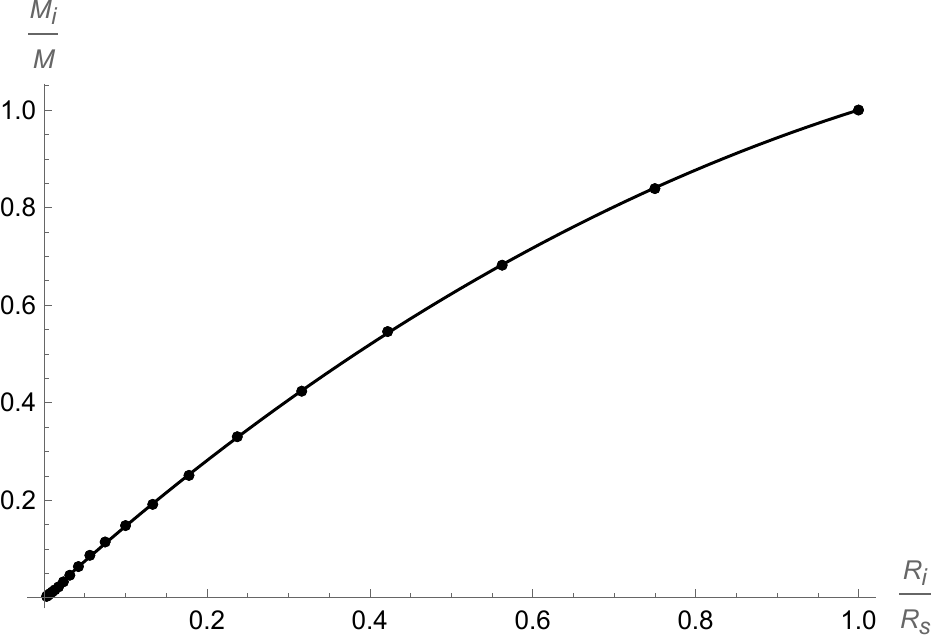}
\end{minipage}
\caption{Left panel: linear mass function $M_i$ for $M=3000\,\mpl$ and $\mu=\mpl/10$ with $N=100$ (red dots)
and $N=80$ (blue dots).
The outer points overlap, while the inner $\Delta N=20$ red dots lie close to the origin as~\eqref{80layers}
and~\eqref{100layers} suggest.
Right panel: quantum corrected mass function for $N=20$ with $M=3000 \, \mpl$ and $\mu=\mpl/10$.}
\label{plot dots}
\end{figure}
As an example, we here consider a refined mass distribution $\mathcal{M}_i$ obtained
 for $N=20$,
$M=3000\,\mpl$ and $\mu=10/\mpl$ following the procedure outlined in Section~\ref{sec2}.
(An alternative method is described in Appendix~\ref{A.1}.)
We first computed $\Delta \mu$ and summed over its columns to determine the mass of each layers,
finally fitted to derive a numerical expression for $\mathcal{M}_i$, which takes the form
\be
\frac{\mathcal{M}_i}{M}
\simeq 
1.53
\,
\frac{R_i}{R_{\rm s}}
-0.533
{\left(
\frac{R_i}{R_{\rm s}}
\right)}^{1.90}
\equiv
a\,x+b\,x^c
\ ,
\label{Mi2}
\ee
where we introduced the dimensionless $x=R_i/R_{\rm s}$. 
The continuous and discrete profiles of $\mathcal{M}_i$ are presented in the right panel of
Fig.~\ref{plot dots}.
The plot reveals a remarkable deviation from the linear profile found within the original model,
reflecting the altered distribution of matter across the layers.
This correction remains valid regardless of the values of $N$, provided the total mass $M$
allows for the existence of a monotonic sequence of $N_i$.
With this in mind, as in the linear approximation of Ref.~\cite{Casadio:2023ymt},
one can exploit $\mathcal{M}_i$ to define an effective continuous MSH mass from
Eq.~\eqref{Mi2}, which can then be substituted into the metric to obtain the line element
\be
\label{gregular}
\frac{\d s^2}{R^2_{\rm s}}
\simeq
-
\left[
1
-
\frac{2\, \gn\, M}{R_{\rm s}}
\left(a+b\,x^{c-1}\right)
\right]
\frac{\d t^2}{R^2_{\rm s}}
+
\left[
1
-
\frac{2\, \gn\, M}{R_{\rm s}}
\left(a+b\,x^{c-1}\right)
\right]^{-1}
\d x^2
+
x^2\,
\d\Omega^2
\ .
\ee
It is remarkable that the quantum correction obtained here does not spoil the regularity
of the metric around the origin that appeared in the linear approximation (which
is consistently reproduced in the limit $b\to 0$ and $a\to 1$).
Notice finally that there is no inner horizon inside the ball since $g_{tt}=g^{rr}$ only
vanishes outside the ball, at a value of $x>1$, as the left panel of Fig.~\ref{gtt} illustrates.
\par
From the above metric, we can compute the effective Einstein tensor
$G^{\mu}_{\ \nu}=8\, \pi \, \gn \, T^{\mu}_{\ \nu}$.
Recalling that $x^1=r$ ($x^0=t$) is a time (space) coordinate inside the horizon, 
one has:
\be
G^1_{\ 1}
=
-8\, \pi\, \gn\, \rho
=
G^0_{\ 0}
=
8\, \pi\, \gn\, p_r
\ee
and
\be
G^2_{\ 2}
=
G^3_{\ 3}
=
8 \,\pi \, \gn \, p_\perp
\ ,
\ee
from which one finds the effective density and radial pressure
\be
\rho(x)
\simeq
-p_r(x)
\simeq
\frac{M\, (a \, x + b \, c \, x^c)}
{4\,  \pi \, R^3_{\rm s} \, x^3}
\ ,
\ee
and the effective tension 
\be
p_{\perp}(x)
\simeq
%\frac{m''(x)}
%{8 \pi \, R^3 \, x}
%=
-\frac{b \, (c-1) \, c \, M \, x^{c-3}}
{8 \pi \, R^3_{\rm s}}
\neq
0
\ .
\ee
The profiles of $\rho(x)$ and $p_{\perp}(x)$ are shown in the right panel of Fig.~\ref{gtt}.
The non vanishing tension is a new feature of the refined model with respect to the linear
approximation, in which $c=1$ and $p_{\perp}(x)=0$.
This result enriches the internal structure of the ball and depends on the introduction
of quantum interactions among particles, that were previously neglected. 
\par
Note that a negative pressure, as found here, is a general feature of approaches
that aim to regularise classical black hole singularities.
Indeed, in such models a negative pressure is required to sustain the regular core that
replaces the singularity (see Refs.~\cite{Maeda:2021jdc,Bambi:2023try} 
for more details and the corresponding literature).
\section{Conclusions and outlook} 
\label{sec-4}
\setcounter{equation}{0}
In Section~\ref{sec-1}, we first overviewed the quantum dust core model of black holes
from Ref.~\cite{Casadio:2023ymt} and highlighted one of its key features, namely the
relation~\eqref{mlinear} which entails a mass that increases linearly with the areal radius.
However, given the wavefunctions~\eqref{psi} and the shape of the corresponding
probability density shown in Fig.~\ref{N=3 & linear fit} (right panel), it is natural
to investigate how the previous scaling gets affected by quantum superpositions
among the wavefunctions.
Indeed the plot suggests that a fraction of particles in one layer may actually be localized
in another one with some non-vanishing probability.
For this reason we refined the model by including these quantum effects,
that modify the value of the mass inside every layers.
\par
The refinemant was described in Section~\ref{sec2}, where we exposed a way to calculate
the amount of mass that each layer gains from the superposition with the others,
along with a discussion on the dependence of the results on the number $N$ of layers.
The method is based on computing the probability that the mass $\mu_i$ of a generic
$i^{\text{th}}$ layer, is actually located in a $j^{\text{th}}$ layer $(j \neq i)$ using the
probability density~\eqref{P_i}.
This is extended for every layer, recovering a new collective mass distribution
$\mathcal{M}_i$ that incorporates the effects of superpositions.
\par
In Section~\ref{sec3}, we provide a numerical evaluation of $\mathcal{M}_i$, for
$N=20$.
Its plot is shown in Fig.~\ref{plot dots} (right panel), which exhibits a deviation from
the linear profile as expected.
Indeed, the fitted function scales approximately like a second order polynomial in the
radial variable with a downward concavity.
The parabolic nature of the mass distributions~\eqref{Mi2} keeps the new effective
metric~\eqref{gregular} regular in the region $0 \leq r \leq R_{\rm s}$ and free from
inner horizons as in Ref.~\cite{Casadio:2023ymt}.
From this metric, we also solved the Einstein field equation to obtain an expression
for $\rho \simeq-p_r$  and $p_{\perp}$ and whose profiles are illustrated in the right panel of
Fig.~\ref{gtt}.
\par
Of course, the shortcomings highlighted in Ref.~\cite{Casadio:2023ymt} are still
present and further adjustments can be made, like improving the fitting accuracy
by studying how the iteration of the corrective mechanism affects~\eqref{Mi2}.
This process may converge to a more precise mass distribution of the ball,
although we leave these analysis for a future work.
In conclusion both the original discussion in Ref.~\cite{Casadio:2023ymt}
and the present refinement, seem to point toward a common conclusion:
quantum effects may regularise the central singularity.
\section*{Acknowledgments}
R.C.~is partially supported by the INFN grant FLAG.
A.G.~is supported in part by the Science and Technology Facilities Council
(grants numbers~ST/T006048/1 and~ST/Y004418/1).
A.M.~is partially supported by MUR under the PRIN2022 PNRR project
n.~P2022P5R22A.
This work has been carried out in the framework of the activities of the
Italian National Group of Mathematical Physics [Gruppo Nazionale per
la Fisica Matematica (GNFM), Istituto Nazionale di Alta Matematica (INdAM)]
\appendix
\section{Alternative formulation}
\label{A.1}
\setcounter{equation}{0}
We briefly present here an alternative method for determining the mass function 
that is equivalent to the one employed in Section~\ref{sec2} but is based on the
spreading of the mass of each layer over the whole ball.
In particular, the mass $\mu_i$ is now weighted by the probability density that
it is exactly confined within $R_i$ and $R_{i+1}$.
This construction is then applied cumulatively for every layer starting from
the innermost all the way to the surface.
\par
Let us call $m_i$ the mass $\mu_i$ weighted by its probability density:
\be
\d{m}_i
=
\mu_i \, {\mathcal{P}}_i\,\d r
=
4\,\pi\,\mu_i \, \abs{\psi_i}^2 \, r^2\,\d r
\ ,
\ee
such that every layer has its own weighted mass.
Then, we can define the cumulative mass $m$ as the sum of all these masses:
\be
\d m
%=\left(\mu_{0}\mathcal{P}_0+...+\mu_{N}\mathcal{P}_N\right)\d r
=
\sum_{i=0}^{N} \d m_i
\ ,
\ee
which is strictly monotonic going from the innermost to the outermost layer.
Finally we define the cumulative mass $\hat{\mathcal{M}}_{j}$
as the following integral over the radius:
\be
\label{M2}
\hat{\mathcal{M}}_{j}
=
\int_{0}^{R_j}\d m
=
4\,\pi\sum_{i=0}^{N}\int_{0}^{R_j} \mu_i \, \abs{\psi_i}^2 \, r^2\,\d r
\, 
\ .
\ee
With this approach we are first assigning a probabilistic mass to
each layer and then summing over them cumulatively.
In a sense this second approach is conceptually more static than
the previous one, where we conceived the particles as free to redistribute
in each layers.
However, it is easy to check that Eq.~\eqref{M2} can be obtained 
from Eq.~\eqref{finalM}, that is
\be
\begin{split}
        \mathcal{M}_j
        \equiv
        &\,
        4\,\pi
        \sum_{k=0}^j\sum_{i=0}^{N}\int_{R_{k}}^{R_{k+1}}\mu_{i}\abs{\psi_{i}}^2
        r^2 \d r
        \quad \quad \quad \forall\ 0\leq i, j\leq N
        \ ,
        \\
        =
        &\,
        4\,\pi
        \sum_{i=0}^{N}\int_{0}^{R_j}\mu_{i} \, \abs{\psi_{i}}^2 r^2 \d r
        \quad \quad \quad \forall\ 0\leq j\leq N
        \ ,
        \\
        =
        &
        \,
        \hat{\mathcal{M}}_j
        \ ,
    \end{split}   
\ee
as it should.
\newpage

\section*{Conflict of Interest}
The authors have no conflicts of interest to declare that are relevant to the content of this article.

\medskip

\bibliographystyle{utphys}
\bibliography{QCbib}
\end{document}